\documentclass[letterpaper, preprint, paper,11pt]{AAS}	
\usepackage{graphicx}
\usepackage{url}
\usepackage{amsmath}
\usepackage{amsfonts}
\usepackage{amssymb}
\usepackage{float}
\usepackage{booktabs}
\usepackage{array}

\usepackage{enumitem}
\usepackage{tikz}
\usepackage{cite}
\usepackage{siunitx}
\usetikzlibrary{arrows.meta,calc,angles,quotes}

\usepackage{algorithm}
\usepackage{algpseudocode}

\usetikzlibrary{arrows.meta, positioning, shapes.geometric}

\newcommand{\bbR}{\mathbb{R}}
\newcommand{\bbN}{\mathbb{N}}
\newcommand{\bbE}{\mathbb{E}}
\newcommand{\bbC}{\mathbb{C}}

\newcommand{\calN}{\mathcal{N}}
\newcommand{\calF}{\mathcal{F}}
\newcommand{\calC}{\mathcal{C}}

\newcommand{\calL}{\mathcal{L}}
\newcommand{\calI}{\mathcal{I}}
\newcommand{\calM}{\mathcal{M}}

\newcommand{\bfc}{\mathbf{c}}

\newcommand{\bfr}{\mathbf{r}}
\newcommand{\bfu}{\mathbf{u}}
\newcommand{\bfv}{\mathbf{v}}
\newcommand{\bfw}{\mathbf{w}}
\newcommand{\bfx}{\mathbf{x}}
\newcommand{\bfz}{\mathbf{z}}

\newcommand{\bfzero}{\mathbf{0}}
\newcommand{\bfone}{\mathbf{1}}

\newcommand{\bfmu}{\boldsymbol{\mu}}

\newcommand{\bfsigma}{\boldsymbol{\sigma}}
\newcommand{\bfpsi}{\boldsymbol{\psi}}

\newcommand{\Key}{\mathsf{KeyGen}}
\newcommand{\Enc}{\mathsf{Enc}}
\newcommand{\Dec}{\mathsf{Dec}}
\newcommand{\sk}{\mathsf{sk}}

\newcommand{\Boot}{\mathsf{Boot}}
\newcommand{\ini}{\mathsf{ini}}
\newcommand{\TCA}{\mathsf{TCA}}
\newcommand{\Cov}{\mathsf{Cov}}
\newcommand{\nom}{\mathsf{nom}}
\newcommand{\ws}{\mathsf{ws}}
\newcommand{\KL}{\mathsf{D}_{\mathsf{KL}}}
\newcommand{\nume}{\mathsf{num}}
\newcommand{\den}{\mathsf{den}}
\newcommand{\pk}{\mathsf{pk}}
\newcommand{\evk}{\mathsf{evk}}

\newcommand{\Pc}{P_c}
\newcommand{\Pcmax}{P_{c,\max}}

\newcommand{\Sigmaz}{\Sigma_\bfz}

\DeclareMathOperator*{\argmin}{argmin}

\newtheorem{thm1}{\bf Theorem}
\newtheorem{prop1}{\bf Proposition}
\newtheorem{lem1}{\bf Lemma}
\newtheorem{asm1}{\bf Assumption}
\newtheorem{defn1}{\bf Definition}
\newtheorem{rem1}{\bf Remark}

\newtheorem{cor1}{\bf Corollary}

\newtheorem{prob1}{\bf Problem}

\newenvironment{lem}{\begin{lem1}}{\hfill$\square$\end{lem1}}

\makeatletter
\renewcommand*{\ALG@name}{Protocol}
\makeatother

\PaperNumber{XX-XXX}

\begin{document}

\title{Encrypted Collision Avoidance Maneuver via Variational Optimization}

\author{Yeongjun Jang\thanks{Ph.D. Student, ASRI, Department of Electrical and Computer Engineering, Seoul National University, Seoul, 08826, South Korea.}, Jihoon Suh\thanks{Ph.D., Anduril Industries, Costa Mesa, CA 92626, USA.}, Kenshiro Oguri\thanks{Assistant Professor, School of Aeronautics and Astronautics, Purdue University, West Lafayette, IN 47907, USA.}, and Takashi Tanaka\thanks{Associate Professor, School of Aeronautics and Astronautics, Purdue University, West Lafayette, IN 47907, USA.}
}

\maketitle{}

\begin{center}
\textbf{Extended Abstract}
\end{center}

\section{Introduction}
\label{sec:intro}

Space collisions can cause severe economic damage and generate debris that threatens other spacecraft \cite{LiouShoo09,AnzmLiou10}.
Therefore, when a close approach is predicted, spacecraft operators perform conjunction analysis to assess collision probability and, if necessary, design a collision avoidance maneuver (CAM) subject to operational constraints such as fuel consumption limits \cite{NASA23, PavaDema25}.
In particular, recent advances in space technology have increased the density of objects in low Earth orbit (LEO) \cite{KessCour78,ESA2026,PardAnse21}, making CAMs increasingly important.

At the same time, as commercial organizations and competing nations are becoming more active in LEO, exact trajectory and operational parameters are often treated as sensitive information since they can reveal the capabilities and mission profiles of an operator.
Hence, rather than directly exchanging such information, commercial operators typically rely on a space situational awareness provider \cite{NASA23}, which screens a space catalog for close approaches and generates conjunction data messages (CDMs). 
A CDM consists of an estimated state and uncertainty covariance of each object at the time of closest approach (TCA), obtained from public tracking data that is of relatively lower fidelity compared to the operator's privately owned estimates. 
Consequently, it can be conservative, leading to unnecessary fuel usage, or optimistic, underestimating collision risk and violating safety requirements.

A natural remedy is to have the operators share their private estimates with a mutually trusted third party and delegate the planning. 
For example, NASA's Conjunction Assessment and Risk Analysis (CARA) service performs risk assessment and provides maneuver recommendations for its mission operations teams \cite{NASA23}.
This approach, however, relies on organizational trust and may be unrealistic for competing operators.
Moreover, even when such a party exists, its service may be too costly, and the party may be semi-honest, meaning that it correctly executes the assigned protocol while attempting to infer sensitive information from the data it receives. 

In this context, there has been a growing interest in privacy-preserving conjunction analysis \cite{KammWill15, HemeLust16, LageHorm25, SuhjMich25}.
Using secure multi-party computation (SMPC) \cite{CramDamg15} and/or homomorphic encryption (HE) \cite{CheoKima17}, these works allow operators to compute the collision probability at TCA without revealing their private information to each other or to the third party. 
However, they only address the assessment step, leaving CAM design unexplored.
The main difficulty is that, unlike conjunction analysis, which reduces to a simple evaluation, standard CAM design is formulated as an optimization problem whose solution typically involves gradient evaluation, comparison, and division \cite{Arme21, DuttMisr22}. 
Since SMPC and HE natively support only addition and multiplication efficiently, these operations are costly and challenging to implement securely.

A similar challenge has drawn interest in the encrypted control literature, where the goal has been to outsource model predictive control to an untrusted server without revealing the client's state or model via HE \cite{AlexMora18, DaruRedd18, SchlIann25, SuhjJang26}.
Since solving the underlying constrained optimization problem typically requires projection or comparison, prior approaches have either delegated performing such operations to the client \cite{AlexMora18, DaruRedd18} or reformulated the problem into a sequence of penalized unconstrained problems involving only polynomial operations, whose real-time practicality remains uncertain \cite{SchlIann25}.
In contrast, our earlier work \cite{SuhjJang26} proposed a variational approach that reformulates the optimization problem as a sampling based estimation problem, which only requires a single round of polynomial computation with no intermediate decryption by the client.

In this paper, we adapt the approach of \cite{SuhjJang26} to develop an encrypted CAM protocol that enables operators to outsource maneuver design to an untrusted server without revealing their private information.
A spacecraft's true state (position/velocity) are sensitive but cannot be measured exactly, and thus, CAM design relies on state estimates and their associated uncertainty covariance.
Since the accuracy of the estimates ultimately depends on the covariance \cite{NekoSkog18,SeriMasi20}, we consider it as the sensitive information, and the state estimates as public.
This allows us to avoid unnecessary encrypted evaluations, keeping the proposed method HE-compatible and computationally feasible within the CAM planning timescale.
To the best of our knowledge, this is the first work to solve a CAM optimization problem under encryption.

The main contributions are summarized as follows. 

\begin{itemize}
    \item We reformulate the standard constrained CAM design optimization problem as a sampling based estimation problem via a variational approach.
    By introducing a soft feasibility weight and a constant density approximation for collision probability, we reduce the private covariance dependent terms to only involve addition and multiplication, making the estimator well suited for HE.
    \item We propose an encrypted CAM protocol that only requires a single round of communication, in which the server evaluates the estimator over ciphertexts without intermediate decryption, thus keeping the private covariance protected throughout the computation.
    \item We apply the protocol on six real-world CDMs, demonstrating improved feasibility and reduced fuel consumption compared with CAMs designed using public covariance.
\end{itemize}

\subsection{Notation}

The sets of real numbers, complex numbers, and positive integers are denoted by $\bbR$, $\bbC$, and $\bbN$. 
For a sequence $v_1,\ldots, v_n$ of scalars or vectors, we define $[v_1;\cdots;v_n]:=[v_1^\top, \cdots, v_n^\top]$.
The Hadamard product (element-wise multiplication) is denoted by $\odot$.
The identity and the zero matrix are denoted by $I$ and $\bfzero$, respectively, with their dimensions indicated as subscripts when necessary.

\section{Preliminaries}

\subsection{Homomorphic Encryption}

Homomorphic encryption (HE) is a cryptographic primitive that enables arithmetic operations to be directly evaluated over encrypted data without decryption \cite{RiveAdle78,Gent09,AcarAksu18}.
Given encryptions of plaintexts $m_1$ and $m_2$, homomorphic addition and multiplication produce ciphertexts that decrypt approximately to $m_1+m_2$ and $m_1m_2$, respectively.
This property allows a user with limited computing power to outsource computationally intensive tasks to an untrusted server, while preserving privacy. 

HE schemes are classified according to the set of operations they support and how many times those operations can be applied without corrupting the underlying plaintext. 
A \textit{partially HE} scheme supports either addition or multiplication for an unlimited number of times, and is accordingly called additively homomorphic (e.g., Paillier \cite{Pail99}) or multiplicatively homomorphic (e.g., RSA \cite{RiveSham78}, ElGamal \cite{Elga85}), respectively.
A \textit{somewhat HE} or \textit{leveled HE} scheme supports both addition and multiplication, but only for a limited number of times. 
Many modern HE schemes that fall into this category are based on the (Ring) Learning With Errors Problem \cite{Rege09,LyubPeik10}, which makes them post-quantum secure unlike \cite{Pail99,RiveSham78,Elga85}, meaning that they are secure even against an adversary equipped with a quantum computer. 
Such schemes inject a small random noise during encryption that makes the decryption approximate, i.e., the decrypted result recovers the correct value only up to some small error.
Under recursive homomorphic operations, the effect of noise accumulates, thus limiting the number of operations that can be performed sequentially.
Thanks to the breakthrough by \cite{Gent09}, any somewhat or leveled HE scheme can be turned into a \textit{fully HE} scheme that supports both addition and multiplication for an unlimited number of times.
The key idea is to periodically apply the bootstrapping technique that homomorphically evaluates the decryption algorithm, thus refreshing the accumulated noise.

\subsection{Multiparty CKKS Cryptosystem} 

We briefly introduce a two-party version \cite{MoucTron21} of the Cheon-Kim-Kim-Song (CKKS) cryptosystem \cite{CheoKima17}, which is a representative fully HE scheme that will be used throughout the paper.

The cryptosystem consists of algorithms $\Key$, $\Enc$, and $\Dec$. 
Let $S_1$ and $S_2$ denote the participating parties.
They first agree on public CKKS paramters, including the ring dimension, modulus chain, and security parameter. 
Through the key generation protocol $\Key$, each party $S_i$ independently generates a secret key share $\sk_i$, and then, jointly generate a public key $\pk$ and an evaluation key $\evk$ required for encryption and performing homomorphic operations, respectively.
Using $\pk$, both operator can encrypt a plaintext $m\in\bbC^{N/2}$ as a ciphertext $\bfc \in \calC$ using the encryption algorithm $\Enc$, where $\calC$ denotes the ciphertext space.
Through a joint protocol $\Dec$, the operators can decrypt $\bfc$ back to a vector $\Dec(\bfc)\in\bbC^{N/2}$ that approximately recovers the plaintext in the sense that $m\approx \Dec(\Enc(m))$, where the approximation error arises from the noise injected during encryption.
Neither secret share $\sk_i$ is disclosed to the other party, and neither party can decrypt $\bfc$ alone.

CKKS supports both addition and multiplication between ciphertexts, as well as multiplication between a plaintext and a ciphertext.
Specifically, there exist homomorphic operations 
\begin{align*}
    \oplus &: \calC \times \calC \to \calC, &
    \otimes &: \calC \times \calC \to \calC, & \otimes_p &: \bbC^{N/2} \times \calC \to \calC,
\end{align*}
such that
\begin{align}\label{eq:homo}
    \Dec(\Enc(m_1) \oplus \Enc(m_2)) &\approx m_1 + m_2, \\
    \Dec(\Enc(m_1) \otimes \Enc(m_2)) &\approx m_1 \odot m_2, \nonumber \\
    \Dec( m_1 \otimes_p \Enc(m_2)) &\approx m_1 \odot m_2, \nonumber
\end{align}
for all $m_1,m_2\in\bbC^{N/2}$.

The encryption noise associated with each ciphertext accumulate when the above homomorphic operations are repeatedly applied without decryption. 
Consequently, without additional treatment, only a limited number of sequential operations can be evaluated.
This limitation can be overcome by employing the bootstrapping operation
\begin{align*}
    \Boot:\calC \to \calC,    
\end{align*} 
that refreshes a ciphertext and resets its computational capacity \cite{CheoHank18}.
However, it is computationally far more costly than homomorphic addition and multiplication, making it the main bottleneck.

If the message to be encrypted is a vector of dimension less than $N/2$ or a scalar, it can be embedded into $\bbC^{N/2}$ by zero-padding.
Then, after decryption, the desired result can be obtained from the corresponding slots of the decrypted vector.
Furthermore, multiple messages can be \textit{packed} into different slots of a single plaintext, allowing homomorphic operations to be performed on them in parallel.
This property is referred to as single-instruction-multiple-data (SIMD), and is essential in improving computational efficiency.

\section{Problem Formulation}
\label{sec:formulation}

\subsection{Spacecraft Dynamics}
\label{subsec:uncertainty}

We consider two spacecraft $S_1$ and $S_2$, each modeled by the two-body gravitational dynamics:
\begin{align}\label{eq:true_state_decomp}
    \frac{d}{dt}\bfx_i = f(\bfx_i,\bfu_i):=
    \begin{bmatrix}
        \bfv_i \\
        -\mu\dfrac{ \bfr_i}{\|\bfr_i\|^3} + \bfu_i   
    \end{bmatrix}, \qquad \bfx_i(0)=\bfx_i^\ini, \qquad i=1,2,
\end{align}
where $\bfx_i:=[\bfr_i;\bfv_i]\in\bbR^6$ is the state with the initial value $\bfx_i^\ini\in\bbR^6$, where $\bfr_i\in\bbR^3$ and $\bfv_i\in\bbR^3$ denote the position and velocity of $S_i$ in the Earth-Centered Inertial (ECI) frame, respectively, $\bfu_i\in\bbR^3$ is the control acceleration, and $\mu>0$ is the Earth's standard gravitational parameter.

Let $\bar{\bfx}_i$ denote the nominal state trajectory of spacecraft $S_i$ that represents an estimation of the unknown true state $\bfx_i$.
The nominal state is governed by the uncontrolled dynamics
\begin{align*}
    \frac{d}{dt}\bar{\bfx}_i =f(\bar{\bfx}_i,0), \qquad \bar{\bfx}_i(0) = \bar{\bfx}_i^\ini, \qquad i=1,2,
\end{align*}
where $\bar{\bfx}_i^\ini\in\bbR^6$ is the prediction of the true initial value $\bfx_i^\ini$.

Define the deviation of the true state from the nominal state as
\begin{align*}
    \delta\bfx_i = 
    \begin{bmatrix}
        \delta \bfr_i\\
        \delta \bfv_i   
    \end{bmatrix}
    := \bfx_i - \bar{\bfx}_i \in\bbR^6.
\end{align*}
Linearizing \eqref{eq:true_state_decomp} about the nominal trajectory $(\bar{\bfx}_i,0)$ yields the following linear time-varying dynamics for the state deviation:
\begin{align}\label{eq:dynLin}
    \frac{d}{dt}\delta\bfx_i = A_i(t) \delta \bfx_i + B \bfu_i,
\end{align}
where 
\begin{align*}
    A_i(t) := 
    \begin{bmatrix}
        0 & I \\ 
        -\theta \dfrac{\partial}{\partial \bfr} \dfrac{\bfr}{\|\bfr\|^3}\bigr|_{\bfr = \bar{\bfr}_i(t)} & 0    
    \end{bmatrix} =
    \begin{bmatrix}
        0 & I \\ 
        -\dfrac{\theta}{\|\bar{\bfr}_i(t)\|^3}I
+
\dfrac{3\theta}{\|\bar{\bfr}_i(t)\|^5}
\bar{\bfr}_i(t)\bar{\bfr}_i(t)^\top & 0    
    \end{bmatrix}
    , \quad 
    B:=\begin{bmatrix}
        0 \\ I
    \end{bmatrix}.
\end{align*}

\subsection{Encounter Geometry and Collision Probability}

We are interested in the conjunction analysis of $S_1$ and $S_2$ at the time of closest approach (TCA), denoted by $t_\TCA\ge 0$, at which the nominal states of the two spacecraft are predicted to be closest.
Due to orbit-determination error, the true initial state may deviate from its nominal estimate.
We model these initial deviations as independent zero-mean Gaussian, and by propagating them through the linearized dynamics \eqref{eq:dynLin}, the position deviation at TCA with zero input, $\delta \bfr_i^\nom(t_\TCA)$, is also obtained as a Gaussian with\footnote{Here, the superscript $\nom$ indicates that these are evaluated in the absence of any collision avoidance maneuver.}
\begin{subequations}\label{eq:init_uncertainty}
\begin{align}
    &\delta\bfr_i^\nom(t_\TCA) \sim \calN(\bfzero,\,P_{i}(t_\TCA)), \qquad i=1,2,
    \label{eq:init_gaussian}\\
    &\Cov\!\left(\delta\bfr_1^\nom(t_\TCA),\,\delta\bfr_2^\nom(t_\TCA)\right) = \bfzero.
    \label{eq:init_indep}
\end{align}
\end{subequations}
Here, $P_{i}(t_\TCA) \in \bbR^{3\times 3}$ denotes the position deviation covariance
of $S_i$ at $t_\TCA$. 
This covariance may be computed from a publicly available, lower-fidelity conjunction data message (CDM), or from higher-fidelity tracking and estimation data privately owned by the spacecraft operator.

To assess collision risk, the relative position of the two spacecraft is projected onto the \emph{encounter plane}, which is defined as the two-dimensional plane orthogonal to their relative velocity at $t_\TCA$.
Formally, let $\Pi\in\bbR^{2\times 3}$ with $\Pi\Pi^\top = I$ be the projection matrix onto that plane. 
Since the position deviations of the two spacecraft are independent Gaussian, the projected relative position 
\begin{align}\label{eq:projRelPos}
    \bfz^\nom := \Pi \left( \left(\bar{\bfr}_2(t_\TCA) - \bar{\bfr}_1(t_\TCA) \right) + \left( \delta \bfr_2^\nom(t_\TCA) - \delta \bfr_1^\nom(t_\TCA) \right) \right) \in \bbR^2,
\end{align}
is also Gaussian, with mean and covariance
\begin{subequations}\label{eq:projMeanCov}
\begin{align}
    \bfmu_\bfz^\nom &:= \bbE \left[ \bfz^\nom \right] = \Pi \left(\bar{\bfr}_2(t_\TCA) - \bar{\bfr}_1(t_\TCA) \right) \label{eq:projMean}
    \\
    \Sigma_\bfz^\nom &:= \Cov(\bfz^\nom,\bfz^\nom) = \Pi (P_1(t_\TCA) + P_2(t_\TCA)) \Pi^\top.\label{eq:projCov}
\end{align}
\end{subequations}

Under the standard short-encounter approximation \cite{FostEste92}, the probability of collision can be computed by integrating the position deviation probability density over the collision region \cite{akella2000probability}.
With $R>0$ denoting the combined hard-body radius of $S_1$ and $S_2$, it is explicitly obtained as
\begin{equation}\label{eq:poc_2d_def}
    \Pc(\bfmu_\bfz^\nom,\Sigma_\bfz^\nom) := \iint_{\|\bfz\|_2 \le R} \dfrac{1}{2\pi|\Sigma_\bfz^\nom|^{1/2}} \exp \left(-\tfrac{1}{2}(\bfz-\bfmu_\bfz^\nom)^\top (\Sigma_\bfz^\nom)^{-1}(\bfz-\bfmu_\bfz^\nom) \right)\,d\bfz^\nom.
\end{equation}

\subsection{Collision Avoidance Maneuver}

When $\Pc(\bfmu_\bfz^\nom,\Sigmaz^\nom)$ exceeds a prescribed safety threshold $P_{c,{\max}}>0$, the spacecraft operators must plan and execute a collision avoidance maneuver (CAM).
In what follows, we formulate CAM design as an optimization problem that minimizes fuel consumption, while ensuring that the collision probability drops below $P_{c,{\max}}$.
The formulation permits both spacecraft to maneuver simultaneously.
The more common setting in which only one object can maneuver, for example, when the secondary object, is naturally included as a special case.

For some $M\in\bbN$, let
\begin{align*}
    0 \le t^{(1)} < \cdots < t^{(M)} < t_\TCA
\end{align*}
denote a set of pre-specified candidate maneuver times.
Accordingly, we model the control acceleration of each spacecraft $S_i$ as a finite sum of Dirac deltas at those candidate times:
\begin{align}\label{eq:dirac}
    \bfu_i(t) = \sum_{k=1}^M \Delta \bfv_i^{(k)} \delta(t-t^{(k)}),
\end{align}
where $\Delta \bfv_i^{(k)}\in\bbR^3$ denotes the impulsive velocity increment at $t^{(k)}$.

To analyze the impact of each impulse, we introduce the state transition matrix $\Phi_i(t,\tau)\in\bbR^{6\times 6}$ associated with the linearized dynamics \eqref{eq:dynLin}, defined as the solution of
\begin{align*}
    \frac{\partial}{\partial t} \Phi_i(t,\tau) &= A_i(t) \Phi_i(t,\tau), \\
    \Phi_i(\tau,\tau) &= I,
\end{align*}
for all $0\le \tau \le t \le t_\TCA$.
From $\Phi_i(t,\tau)$, we extract the velocity-to-position transition matrix as
\begin{align*}
    \Gamma_i(t,\tau) :=
    \begin{bmatrix}
        I_3 & \bfzero_{3\times 3}
    \end{bmatrix}
    \Phi_i(t,\tau)
    \begin{bmatrix}
        \bfzero_{3\times 3} \\
        I_3
    \end{bmatrix}.
\end{align*}
Then, the impulse $\Delta \bfv_i^{(k)}$ applied at $t^{(k)}$ produces a position deviation $\Gamma_i(t_\TCA,t^{(k)})\,\Delta \bfv_i^{(k)}$ at TCA.

Since each impulse is deterministic, the entire maneuver \eqref{eq:dirac} only shifts the mean of the position deviation $\delta \bfr_i(t_\TCA)$, resulting in
\begin{align*}
    \delta \bfr_i (t_\TCA) \sim \calN\left(\sum_{k=1}^M \Gamma_i(t_\TCA,t^{(k)}) \, \Delta \bfv_i^{(k)}, P_i(t_\TCA) \right), \qquad i=1,2.
\end{align*}
Then, the projected relative position in the presence of maneuver
\begin{align}
    \bfz := \Pi \left( \left(\bar{\bfr}_2(t_\TCA) - \bar{\bfr}_1(t_\TCA) \right) + \left( \delta \bfr_2(t_\TCA) - \delta \bfr_1(t_\TCA) \right) \right) \in \bbR^2,
\end{align}
remains Gaussian, with its mean shifted and its covariance unchanged relative to  \eqref{eq:projMeanCov}:
\begin{align*}
    \bfmu_\bfz &:= \bbE\left[ \bfz \right] = \bfmu_\bfz^\nom + \Pi \sum_{k=1}^M \left(\Gamma_2(t_\TCA,t^{(k)}) \, \Delta \bfv_2^{(k)} - \Gamma_1(t_\TCA,t^{(k)}) \, \Delta \bfv_1^{(k)}  \right), \\
    \Sigma_\bfz &:= \Cov(\bfz, \bfz) = \Sigma_\bfz^\nom. 
\end{align*}

By defining
\begin{align*}
    \Gamma :=
    \Pi
    \begin{bmatrix}
        -\Gamma_1(t_\TCA,t^{(1)}) & \cdots & -\Gamma_1(t_\TCA,t^{(M)}) & \Gamma_2(t_\TCA,t^{(1)}) & \cdots & \Gamma_2(t_\TCA,t^{(M)})
    \end{bmatrix} \in \bbR^{2\times 6M},
\end{align*}
we compactly rewrite $\bfmu_\bfz$ as
\begin{align*}
    \bfmu_\bfz = \bfmu_\bfz^\nom + \Gamma \Delta \bfv,
\end{align*}
where $\Delta \bfv := [\Delta \bfv_1^{(1)};\cdots;
        \Delta \bfv_1^{(M)};
        \Delta \bfv_2^{(1)};
        \cdots;
        \Delta \bfv_2^{(M)}]\in\bbR^{6M}$
is the stacked decision variable.
Consequently, we formulate the CAM design as the following optimization problem:
\begin{align}
    \min_{\Delta \bfv \in \bbR^{6M}} \quad & \left\| \Delta \bfv \right\|_2^2  \label{eq:prob}\\
    \text{s.t.} \quad & 
    \Pc\left(\bfmu_\bfz, \Sigma_\bfz \right) = \Pc\left(\bfmu_\bfz^\nom + \Gamma \Delta \bfv, \Sigma_\bfz^\nom \right) \le \Pcmax. \nonumber
\end{align}
The rationale is to minimize the total maneuver energy, thus minimizing fuel consumption, subject to keeping the collision probability below $\Pcmax$.
Note that $\bfmu_\bfz^\nom$ and $\Gamma$ depend only on the nominal trajectory and the system model, and hence, can be precomputed.

\subsection{Threat Model and Problem Statement}

The quality of the maneuver obtained from \eqref{eq:prob} depends critically on the fidelity of the covariance $P_i(t_\TCA)$ used to compute $\Sigma_\bfz^\nom$.
A CAM designed using the low-fidelity covariance provided in the CDM may be overly conservative, leading to unnecessary fuel consumption, or overly optimistic, failing to avoid collision.

It is therefore desirable to exploit the operators' privately owned high-fidelity covariance.
However, such information may reveal sensitive intrinsic properties of the spacecraft, and operators may therefore be unwilling to disclose it.

Motivated by this setting, we aim to develop a privacy-preserving protocol in which each operator encrypts its private covariance and sends it to an untrusted third party.
The third party then solves \eqref{eq:prob} directly over ciphertexts and returns the encrypted maneuver, so that neither the other operator nor the third party learns the private covariance. 
Accordingly, for the remainder of the paper, $P_i(t_\TCA)$ denotes the private covariance of spacecraft $S_i$.
In contrast, the quantities $\mu_\bfz^\nom$ and $\Gamma$, the candidate maneuver times $\{t^{(k)}\}_{k=1}^M$, the combined hard-body radius $R$, and the safety threshold $\Pcmax$ are assumed to be public.

\section{Encrypted CAM Design}

\subsection{Limitation of Conventional Gradient-Based CAM Design}

The conventional approach to CAM design is to employ standard gradient based solvers \cite{PatePete03,DuttMisr23}.
Although such methods are widely used, we demonstrate that they are not readily compatible with HE in our setting.

For example, consider the augmented Lagrangian (AL) method \cite{Rock74}, which handles the nonconvex inequality constraint in \eqref{eq:prob} as follows.
Define
\begin{subequations}\label{eq:AL}
\begin{align}
    h(\Delta \bfv) := \Pc\left(\bfmu_\bfz^\nom + \Gamma \Delta \bfv, \Sigma_\bfz^\nom\right) - \Pcmax,
\end{align}
so that the inequality constraint in \eqref{eq:prob} is equivalent to $h(\Delta \bfv) \le 0$.
The augmented Lagrangian is then given by
\begin{align}
    \calL_\rho(\Delta \bfv, \lambda) := \left\| \Delta \bfv \right\|_2^2 + \frac{1}{2\rho} \left( \left[ \lambda + \rho h(\Delta \bfv) \right]_+^2 - \lambda^2 \right),
\end{align}
where $\lambda \ge 0$ is the Lagrange multiplier, $\rho > 0$ is the penalty factor, and $[\cdot]_+ := \max\{\cdot, 0\}$ is the ReLU function.
Let $\Delta \bfv^{(l)}$ and $\lambda^{(l)}$ denote the primal and multiplier estimates at iteration $l = 0, 1, \ldots$, respectively.
At each iteration, the primal variable is updated by solving the unconstrained subproblem
\begin{align}\label{eq:primalUpdate}
    \Delta \bfv^{(l+1)} = \argmin_{\Delta \bfv \in \bbR^{6M}} \calL_\rho(\Delta \bfv, \lambda^{(l)})
\end{align}
using gradient descent, followed by the multiplier update
\begin{align}
    \lambda^{(l+1)} = \left[ \lambda^{(l)} + \rho h(\Delta \bfv^{(l+1)}) \right]_+.
\end{align}
\end{subequations}

The gradient of the augmented Lagrangian required for \eqref{eq:primalUpdate} can be explicitly computed as 
\begin{align}\label{eq:alGrad}
    \nabla_{\Delta \bfv} \calL_\rho(\Delta \bfv, \lambda^{(l)}) &= 2\Delta \bfv + \left[ \lambda^{(l)} + \rho h(\Delta \bfv) \right]_+ \cdot \nabla_{\Delta \bfv} h(\Delta \bfv) \nonumber \\
    &= 2\Delta \bfv + \left[ \lambda^{(l)} + \rho h(\Delta \bfv) \right]_+ \cdot \Gamma^\top \nabla_{\bfmu_\bfz} P_c(\bfmu_\bfz^\nom + \Gamma \Delta \bfv, \Sigma_\bfz^\nom).
\end{align} 
Since $\Sigma_\bfz^\nom$ defined in \eqref{eq:projCov} depends on the private covariances $P_i(t_\TCA)$, the gradient of $P_c$ in the last term of \eqref{eq:alGrad} cannot be computed from public information.
Indeed, one could instead use the public covariance provided in the CDM, but this is undesirable because this can degrade maneuver performance as discussed earlier.
Moreover, the ReLU function appears in both the primal gradient and the multiplier update, requiring comparison operation that is difficult perform efficiently under HE.
These features hinder the direct integration of HE to the AL method \eqref{eq:AL}.

\subsection{Gradient-Free Variational Approach}

Motivated by the limitations of gradient based methods, we reformulate the CAM design problem as a sampling based estimation problem that avoids explicit gradient evaluation. 

Define the feasible set by $\calF := \left\{ \Delta \bfv \in \bbR^{6M} \mid h(\Delta \bfv) \le 0 \right\}$.
Then, the constrained problem \eqref{eq:prob} can be equivalently rewritten as the unconstrained problem 
\begin{align}\label{eq:probEq}
    \min_{\Delta \bfv \in \bbR^{6M}} \phi(\Delta \bfv ), 
\end{align}
where
\begin{align}
    \phi(\Delta \bfv) &:= \| \Delta \bfv \|_2^2 + \calI_\calF(\Delta \bfv), \\
    \calI_\calF(\Delta \bfv) &:= 
    \begin{cases}
        0, & \Delta \bfv \in \calF, \\
        \infty, & \mbox{otherwise}.
    \end{cases} \label{eq:indicator}
\end{align}

Rather than optimizing over a deterministic maneuver $\Delta \bfv$, we seek for an optimal probability distribution $\kappa$ on $\Delta \bfv$.
Given a reference distribution $\kappa_0$, we consider the Kullback-Leibler (KL) regularized problem 
\begin{align}\label{eq:probKL}
    J^*=\min_{\kappa} \bbE_{\Delta \bfv \sim \kappa} \left[ \phi(\Delta \bfv) \right] +  \tau \KL(\kappa \| \kappa_0),
\end{align}
where $\tau >0$ is a temperature parameter and $\KL(\kappa \| \kappa_0)$ is the KL divergence of $\kappa$ from $\kappa_0$.
By the variational formula \cite{BoueMich98} , the optimal distribution admits a closed-form representation obtained by exponentially tilting $\kappa_0$, which importantly, does not require any gradient evaluation.

\begin{lem}\upshape
    For any $\tau>0$, assume that $0<\bbE_{\Delta \bfv \sim \kappa_0} [\exp(-\phi(\Delta \bfv)/\tau)] < \infty$.
    Then, the optimal value of \eqref{eq:probKL} is 
    \begin{align*}
        J^* = -\tau \log \bbE_{\Delta \bfv \sim \kappa_0} \left[ \exp\left(- \frac{\phi(\Delta \bfv)} {\tau} \right) \right]   
    \end{align*}
    and the minimizer is uniquely obtained as
    \begin{align}\label{eq:optDist}
        \kappa^*(\Delta \bfv) = \frac{ \kappa_0(\Delta \bfv) \exp\left(- \frac{\phi(\Delta \bfv)} {\tau} \right) }{\bbE_{\Delta \bfv \sim \kappa_0} \left[ \exp\left(- \frac{\phi(\Delta \bfv)} {\tau} \right) \right]}.
    \end{align}
\end{lem}

We take the mean of the optimal distribution \eqref{eq:optDist} as a point estimate of the maneuver: 
\begin{align}\label{eq:hatv}
    \widehat{\Delta \bfv} := \bbE_{\Delta \bfv \sim \kappa^*}\left[ \Delta \bfv \right] = 
    \frac{ \bbE_{\Delta \bfv \sim \kappa_0} \left[ \Delta \bfv \exp\left(- \frac{\phi(\Delta \bfv)} {\tau} \right)  \right] }{\bbE_{\Delta \bfv \sim \kappa_0} \left[ \exp\left(- \frac{\phi(\Delta \bfv)} {\tau} \right) \right]}.
\end{align}
Since the expectations in \eqref{eq:hatv} are not readily computable under HE, we draw $K\in\bbN$ number of i.i.d. samples $\bfv^{(j)}$ for $j=1,2,\ldots, K$ and approximate it with the sampling based estimate 
\begin{align}\label{eq:hatvSample}
    \widehat{\Delta \bfv}_K := \frac{\sum_{j=1}^K \Delta \bfv^{(j)} \exp\left(- \frac{\phi(\Delta \bfv^{(j)})} {\tau} \right) }{\sum_{j=1}^K \exp\left(- \frac{\phi(\Delta \bfv^{(j)})} {\tau} \right)}.
\end{align}

Here, we choose the reference distribution as a Gaussian distribution 
\begin{align}\label{eq:ws}
    \kappa_0=\calN(\Delta \bfv_{\ws},\Sigma_\ws),
\end{align}
where $\Delta \bfv_\ws$ is a warm start obtained from a single AL iteration using the public covariance provided by CDM, and $\Sigma_\ws$ is a public covariance parameter. 
Since both the mean and the covariance are independent of the private covariance, $\kappa_0$ and the samples drawn from it are treated as public.

Although the sampling based estimate \eqref{eq:hatvSample} avoids gradient evaluation and replaces the expectations in \eqref{eq:hatv} with finite summations, its exponential term still requires evaluating the indicator function \eqref{eq:indicator}, which involves a comparison operation that is not readily supported by HE. 
In the following subsection, we replace the indicator function with a soft feasibility weight and adopt the constant-density approximation of the collision probability, thereby reducing the encrypted computation to the evaluation of a single quadratic form.

\subsection{Soft Feasibility Weight and Constant-Density Approximation of $P_c$}

Let us define the soft violation score by
\begin{align}\label{eq:score}
    s(\Delta \bfv) := \left[ \log \frac{P_c(\bfmu_\bfz^\nom + \Gamma \Delta \bfv, \Sigma_\bfz^\nom)}{\Pcmax} \right]_+ = \left[ \log \frac{h(\Delta \bfv) + \Pcmax}{\Pcmax} \right]_+.
\end{align}
By construction, $s(\Delta \bfv) =0$ whenever $\Delta \bfv \in \calF$, whereas
$s(\Delta \bfv) >0$ when $\Delta \bfv \notin \calF$ and also increases monotonically with constraint violation. 
With this, we define the soft feasibility weight 
\begin{align*}
    w_\eta(\Delta \bfv) := \exp(-\eta s(\Delta \bfv)),
\end{align*}
where the parameter $\eta>0$ determines the sharpness. 
It provides a soft approximation of the feasibility indicator function in the sense that
\begin{align*}
    w_\eta(\Delta \bfv) \xrightarrow{\eta \to \infty} \bfone_\calF (\Delta \bfv) := 
    \begin{cases}
        1, & \Delta \bfv \in \calF, \\
        0, & \mbox{otherwise}.
    \end{cases}
\end{align*}

Next, we adopt the constant-density approximation of the collision probability, which is standard in conjunction analysis \cite{AlfrAkel99}. 
Specifically, we approximate the probability density over the combined hard-body region by its value at the origin. 
Then, 
\begin{align*}
    P_c(\bfmu_\bfz^\nom + \Gamma \Delta \bfv, \Sigma_\bfz^\nom)
    &\approx 
    \tilde{P}_c(\bfmu_\bfz^\nom + \Gamma \Delta \bfv, \Sigma_\bfz^\nom) \\
    &:= 
    \iint_{\|\bfz\|_2 \le R} \dfrac{1}{2\pi|\Sigma_\bfz^\nom|^{1/2}} \exp \left(-\frac{1}{2} \calM^2(\Delta \bfv) \right)\,d\bfz\\
    &= \dfrac{R^2}{2|\Sigma_\bfz^\nom|^{1/2}}
    \exp \left( -\frac{1}{2} \calM^2(\Delta \bfv) \right),
\end{align*}
where $\calM^2(\Delta \bfv):=(\bfmu_\bfz^\nom + \Gamma \Delta \bfv)^\top (\Sigma_\bfz^\nom)^{-1}(\bfmu_\bfz^\nom + \Gamma \Delta \bfv)$ is the squared Mahalanobis distance.

By substituting $\tilde{P}_c$ in place of $P_c$, we obtain 
\begin{align}\label{eq:tildew}
    w_\eta(\Delta \bfv) \approx \tilde{w}_\eta(\Delta \bfv) &:= \exp \left( -\eta \left[ \log \frac{ \tilde{P}_c(\bfmu_\bfz^\nom + \Gamma \Delta \bfv, \Sigma_\bfz^\nom)}{\Pcmax} \right]_+ \right) \nonumber \\
    &= 
    \exp \left( -\frac{\eta}{2} \left[  \left( \calM_{\min}^2 - \calM^2(\Delta \bfv) \right)
    \right]_+ \right),
\end{align}
where 
\begin{align*}
    \calM_{\min}^2 := 2\log \left( \frac{R^2}{2 \Pcmax \left| \Sigma_\bfz^\nom \right|^{1/2}} \right).
\end{align*}
Furthermore, by replacing $\bfone_\calF(\Delta \bfv)$ with $\tilde{w}_\eta(\Delta \bfv)$, we can approximate the sampling based estimate \eqref{eq:hatvSample} as
\begin{align}\label{eq:tildev}
    \widehat{\Delta \bfv}_K &= \frac{\sum_{j=1}^K \Delta \bfv^{(j)} \exp\left(- \frac{ \left\| \Delta \bfv ^{(j)} \right\|_2^2} {\tau} \right) \bfone_\calF(\Delta \bfv^{(j)}) }{\sum_{j=1}^K \exp\left(- \frac{\left\| \Delta \bfv ^{(j)} \right\|_2^2} {\tau} \right) \bfone_\calF(\Delta \bfv^{(j)})} \nonumber \\
    &\approx  \frac{\sum_{j=1}^K \Delta \bfv^{(j)} \exp\left(- \frac{ \left\| \Delta \bfv ^{(j)} \right\|_2^2} {\tau} \right) \tilde{w}_\eta(\Delta \bfv^{(j)}) }{\sum_{j=1}^K \exp\left(- \frac{\left\| \Delta \bfv ^{(j)} \right\|_2^2} {\tau} \right) \tilde{w}_\eta(\Delta \bfv^{(j)})} =: \widetilde{\Delta \bfv}_K.
\end{align}

\begin{algorithm}[t]
\caption{Encrypted Variational CAM}
\label{alg:enc}
\begin{algorithmic}[1]
\small
\algrenewcommand\algorithmicindent{0.9em}

\renewcommand{\algorithmicrequire}{\# \textbf{Offline procedure (operators)}}
\Require

\State Compute $\bfsigma$ and $\calM_{\min}^2$ from the private covariance
\State Encrypt and send
$\bigl(\{\Enc(\bfsigma_r)\}_{r=1}^3,\Enc(\calM_{\min}^2)\bigr)$
to the server

\renewcommand{\algorithmicrequire}{\# \textbf{Offline procedure (server)}}
\Require

\State $\Delta\bfv_{\ws}\gets$ One AL iteration using the public CDM covariance
\State $\kappa_0\gets\calN(\Delta\bfv_{\ws},\Sigma_{\ws})$
\For{$j=1,\ldots,K$}
    \State Sample $\Delta\bfv^{(j)}\sim\kappa_0$
    \State $\bfmu_\bfz^{(j)}
    \gets
    \bfmu_\bfz^\nom+\Gamma\Delta\bfv^{(j)}$
    \State Compute $\bfpsi^{(j)}$ and $\exp\left(-\frac{\|\Delta \bfv^{(j)}\|_2^2}{\tau}\right)$
\EndFor

\renewcommand{\algorithmicrequire}{\# \textbf{Encrypted evaluation (server)}}
\Require

\For{$j=1,\ldots,K$}
    \State $\bfc^{(j)} \gets \Enc(\calM_{\min}^2) \oplus \sum_{r=1}^3 (-\bfpsi^{(j)}_r) \otimes_p \Enc(\bfsigma_r)$
    \State $\tilde{w}_\eta(\Delta \bfv)) \gets g(\calM_{\min}^2 - \calM^2(\Delta \bfv^{(j)}))$
\EndFor

\State $\bfc_{\nume} \gets \sum_{j=1}^K \Delta \bfv^{(j)} \exp\left(- \frac{ \left\| \Delta \bfv ^{(j)} \right\|_2^2} {\tau} \right) \otimes_p \tilde{\bfw}^{(j)} $

\State $\bfc_{\den} \gets \sum_{j=1}^K  \exp\left(- \frac{ \left\| \Delta \bfv ^{(j)} \right\|_2^2} {\tau} \right) \otimes_p \tilde{\bfw}^{(j)}$

\State Return $(\bfc_{\nume},\bfc_{\den})$ to the operators

\renewcommand{\algorithmicrequire}{\# \textbf{Finalization (operators)}}
\Require

\State $\widetilde{\Delta\bfv}_{K,\Dec} \gets  \frac{\Dec(\bfc_{\nume})}{\Dec(\bfc_{\den})}$
\State Set maneuver components below a prescribed threshold to zero

\end{algorithmic}
\end{algorithm}

\subsection{Encrypted Protocol}

The proposed \textit{encrypted CAM} protocol is summarized in Protocol~\ref{alg:enc}. 
The main idea is to factorize the estimate $\widetilde{\Delta \bfv}_K$ in \eqref{eq:tildev} into public and private components, so that only the latter are evaluated over ciphertexts.
The resulting protocol requires a single round of communication and enables the server to compute an encryption of $\widetilde{\Delta \bfv}_K$ without learning the private covariance. 

The protocol is described in detail as well.
Let
\begin{align*}
    (\Sigma_\bfz^\nom)^{-1} =: 
    \begin{bmatrix}
        \sigma_{11} & \sigma_{12} \\
        \sigma_{12} & \sigma_{22}
    \end{bmatrix}, 
    \qquad 
    \bfsigma := 
    \begin{bmatrix}
        \sigma_{11} \\
        \sigma_{12} \\
        \sigma_{22}
    \end{bmatrix}.
\end{align*}
Since $\Sigma_\bfz^\nom$ is determined by the private covariance $P_i(t_\TCA)$, the vector $\bfsigma$ is also private.
For each sample $\Delta \bfv^{(j)}$, define 
\begin{align}\label{eq:muVec}
    \bfmu_\bfz^{(j)} = \bfmu_\bfz^\nom+\Gamma\Delta\bfv^{(j)}
    =:
    \begin{bmatrix}
        \mu_1^{(j)}\\
        \mu_2^{(j)}
    \end{bmatrix}, \qquad j=1,\ldots, K.
\end{align}
Since $\bfmu_\bfz^\nom$, $\Gamma$, and $\Delta \bfv^{(j)}$ are public, $\bfmu_\bfz^{(j)}$ is also public. 
Hence, the corresponding squared Mahalanobis distance $\calM^2(\Delta \bfv^{(j)})$ can be expressed as the inner product of a vector consisting only of public information and the private vector 
$\bfsigma$:
\begin{align}\label{eq:mahal_factor}
    \calM^2(\Delta \bfv^{(j)})
    =
    (\bfpsi^{(j)})^{\top} \bfsigma ,
\end{align}
where
\begin{align}\label{eq:psi}
    \bfpsi^{(j)}
    :=
    \begin{bmatrix}
        (\mu_1^{(j)})^2\\
        2\mu_1^{(j)}\mu_2^{(j)}\\
        (\mu_2^{(j)})^2
    \end{bmatrix}, \qquad j=1,\ldots, K.
\end{align}
Thus, although $\calM^2$ is quadratic in $\bfmu_\bfz^{(j)}$, it is linear in the private vector $\bfsigma$.

During the offline procedure, the operators compute the private quantities $\bfsigma$ and $\calM_{\min}^2$, encrypt them, and send $( \{\Enc(\bfsigma_r)\}_{r=1}^3,\Enc(\calM_{\min}^2))$ to the server. 
Here, $\bfsigma_r$ denotes the $r$-th component of $\bfsigma$ that are separately encrypted.
In parallel, the server constructs the reference distribution $\kappa_0$ as in \eqref{eq:ws} by performing one AL iteration based on the public CDM covariance. 
It then draws $K$ number of i.i.d. samples $\Delta \bfv^{(j)}\sim \kappa_0$ and computes $\bfpsi^{(j)}$ in \eqref{eq:psi} and $\exp(-\|\Delta \bfv^{(j)}\|_2^2/\tau)$ in plaintext.

After receiving $( \{\Enc(\bfsigma_r)\}_{r=1}^3,\Enc(\calM_{\min}^2))$ the server evaluates 
\begin{align*}
    \bfc^{(j)} = \Enc(\calM_{\min}^2) \oplus \sum_{r=1}^3 (-\bfpsi^{(j)}_r) \otimes_p \Enc(\bfsigma_r),
\end{align*}
where $\bfpsi^{(j)}_r$ likewise denotes the $r$-th component of $\bfpsi^{(j)}$, and the summation is taken with respect to the operation $\oplus$.
Thus, $\bfc^{(j)}$ is an encryption of $\calM_{\min}^2 - \calM^2(\Delta \bfv^{(j)})$.

To obtain an encryption of the soft feasibility weight $\tilde{w}_\eta(\Delta \bfv))$ in \eqref{eq:tildew}, we define the function $g(\cdot) := \exp(-\frac{\eta}{2} [\cdot]_+)$, so that 
\begin{align}\label{eq:g}
    \tilde{w}_\eta(\Delta \bfv)) = g(\calM_{\min}^2 - \calM^2(\Delta \bfv^{(j)})).
\end{align}
Since $g$ involves both the exponential and ReLU functions, which are not efficiently supported by HE, we approximate it by a degree-$l$ Chebyshev polynomial $g_l$ over a prescribed interval $[-\underline{B},\overline{B}]$ \cite{KhanMich23}.  
Because $g_l$ consists of polynomial operations, it can be homomorphically evaluated by composing the operations in \eqref{eq:homo}.
With a slight abuse of notation, we write $g_l(\bfc)$ for the homomorphic evaluation of $g_l$ on a ciphertext $\bfc$ with the polynomial coefficients treated as plaintexts.
Accordingly, the server computes
\begin{align*}
    \tilde{\bfw}^{(j)} = g_l( \bfc^{(j)}),
\end{align*}
which approximates an encryption of $\tilde{w}_\eta(\Delta \bfv)) $.
In fact, for fixed $B$, the approximation error over this interval converges to zero as $l\to \infty$.

Finally, the numerator and denominator of $\widetilde{\Delta \bfv}_K$ are evaluated separately as 
\begin{align*}
    \bfc_{\nume} &= \sum_{j=1}^K \Delta \bfv^{(j)} \exp\left(- \frac{ \left\| \Delta \bfv ^{(j)} \right\|_2^2} {\tau} \right) \otimes_p \tilde{\bfw}^{(j)} \\
    \bfc_{\den} &= \sum_{j=1}^K  \exp\left(- \frac{ \left\| \Delta \bfv ^{(j)} \right\|_2^2} {\tau} \right) \otimes_p \tilde{\bfw}^{(j)}.
\end{align*}
The server returns $(\bfc_\nume, \bfc_\den)$ to the operators, who decrypt them to form 
\begin{align*}
    \widetilde{\Delta\bfv}_{K,\Dec} :=  \frac{\Dec(\bfc_{\nume})}{\Dec(\bfc_{\den})} \approx \widetilde{\Delta\bfv}_K.
\end{align*}
The discrepancy arises from the approximate arithmetic of CKKS and the polynomial approximation of $g$. 
Subsequently, components that are below a prescribed threshold are set to zero before the maneuver is executed.

\section{Numerical Results}
\label{sec:results}

We evaluate the proposed protocol on six real-world CDMs provided by Privateer Space, each with a TCA of May 14, 2026.
The scenarios span miss distances from $77$--\SI{243}{\meter} and relative speeds from $8.37$--\SI{14.41}{\km/\s}.
Full details of the CDMs for all six conjunctions, including object names, collision probabilities, and screen periods, are fully available in our code \url{https://github.com/yj-jang-98/ENCCAM}.

For each case, we used $M=30$ candidate maneuver times, so that $\Delta v\in\mathbb{R}^{180}$. 
Because the operators' private covariances are unavailable, we constructed a surrogate private covariance for each conjunction by rotating the public CDM covariance by $25^\circ$ and scaling it by $0.83$.
We used the public covariance to obtain the warm start for the reference sampling distribution, and the surrogate private covariance for the encrypted protocol.

We drew $K=1024$ samples from (20), and set $\Sigma_{\ws}=I$, $\eta=100$, and $\Pcmax = 10^{-4}$. 
The function $g$ in \eqref{eq:g} was approximated by a degree $l=13$ Chebyshev polynomial over the fixed interval $[-\underline{B},\overline{B}] =[-80,12]$.
The CKKS parameters were chosen for numerical demonstration and further study is needed to assess the protocol at standard security levels.

\begin{figure}[t]
\centering
\includegraphics[width=\columnwidth]{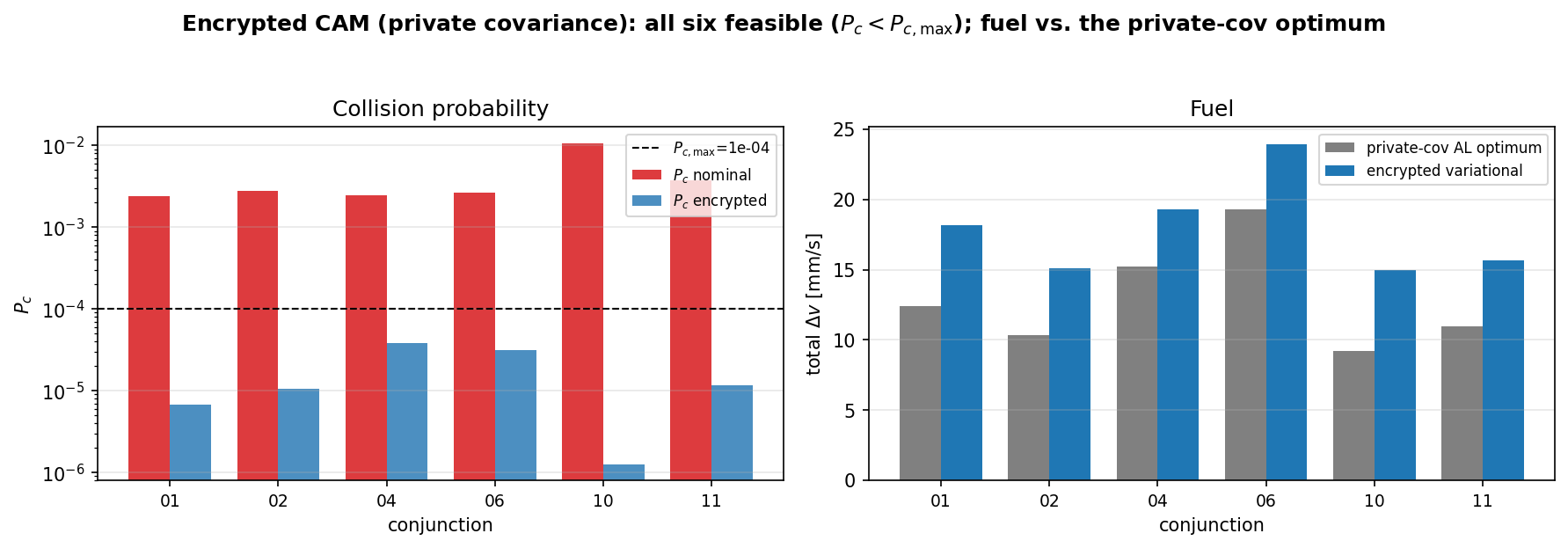}
\caption{Encrypted CAM across the six conjunctions (private covariance).  Left:
the nominal collision probability (red) exceeds $\Pcmax$ for every conjunction,
while the encrypted maneuver (blue) brings it below.  Right: total $\Delta v$ of
the encrypted variational maneuver versus the private-covariance augmented
Lagrangian optimum.}
\label{fig:feas}
\end{figure}

Figure~\ref{fig:feas} compares collision probabilities and total $\Delta \bfv$ across the six conjunctions. 
The nominal collision probability computed using the public covariance exceeds $\Pcmax$ in every case, whereas each
encrypted maneuver brings the collision probability below $\Pcmax$. 
The encrypted maneuvers require more fuel ($24$--$63\%$ more according to Table~\ref{tab:results}) than the baseline AL method designed with the private covariance. 
This difference may arise due to approximations introduced by the variational reformulation and encryption errors.

\section{Future Work for the Full Paper Submission}

For the full paper, we will strengthen the numerical results by performing the following ablation studies.
\begin{itemize}
    \item Vary the rotation and scale of the surrogate private covariance 
    \item Vary the number of samples $K$, parameter $\eta$, Chebyshev polynomial degree $l$
    \item Compare the proposed encrypted protocol with its plaintext counterpart, to quantify the effect of encryption errors
    \item Report runtimes across parameter settings 
\end{itemize}

\section{Acknowledgment}
The first author's effort for this work was supported by the National Research Foundation of Korea (NRF) grant funded by the Korea government (MSIT) (RS-2024-00353032 and RS-2026-25504174).

\bibliographystyle{AAS_publication}   
\bibliography{references}   

\end{document}